# Eighty Years of Neutrino Physics

D. P. Roy
Homi Bhabha Centre for Science Education,
Tata Institute of Fundamental Research, Mumbai 400088, India

## Abstract

This is a pedagogical overview of neutrino physics from the invention of neutrino by Pauli in 1930 to the precise measurement of neutrino mass and mixing parameters via neutrino oscillation experiments in recent years. I have tried to pitch it at the level of undergraduate students, occasionally cutting corners to avoid the use of advanced mathematical tools. I hope it will be useful in introducing this exciting field to a broad group of young physicists.

## 1. Introduction

Neutrino physics originated from the study of radioactive beta decay,

$$N(A,Z) \to N'(A, Z \pm 1) + e^{\mp}. \qquad (1)$$

Thanks to Becquerel and the Curies, several radioactive decays of this type had been investigated by the 1920s. Denoting the momenta of the decay products in the rest frame of the parent nucleus as $\pm p$, one sees that the kinetic energy of N' and the total energy of electron are

$$E_{N'} = p^2 / 2M_{N'} \ \& \ E_e = \sqrt{p^2 + m_e^2}, \qquad (2)$$

where $p \sim$ MeV in natural units ($\hbar = c = 1$). Thus $E_{N'} \ll E_e$, i.e. the recoil energy of the daughter nucleus is negligible relative to the electron energy. Thus by energy conservation one would expect the outgoing electron energy to equal the difference between the parent and daughter nuclear masses ( known as the reaction energy release Q ), i.e.

$$E_e \approx M_N - M_{N'} = Q. \qquad (3)$$

So one expected a monoenergetic line spectrum for the electron. Instead it showed up as a typical continuum spectrum spanning the range $m_e$ – Q, as illustrated in fig.1. This figure is adopted from ref.[1], which we closely follow in this section. To explain this discrepancy Pauli suggested the RHS of the decay (1) to contain a neutral particle, i.e.

$$N(A,Z) \to N'(A, Z \pm 1) + e^{\mp} + \bar{\nu}(\nu), \qquad (4)$$

where the bar on top indicates antiparticle.

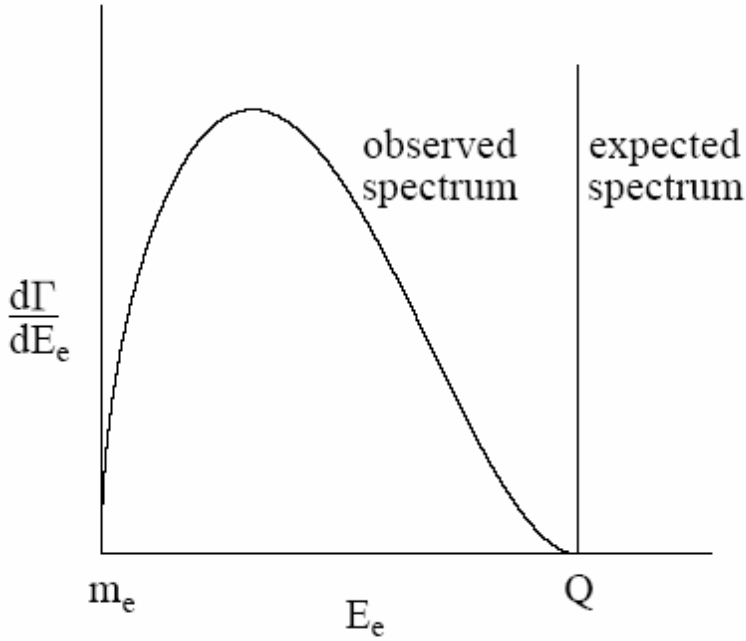

Fig1. The electron energy spectrum in nuclear beta decay [1].

Pauli made this suggestion in an informal letter dated December 1930, where he called it 'neutron'. But by the time he published it in 1933 [2], the neutron had been discovered by Chadwick as the neutral partner of the proton. So it was renamed by Fermi as 'neutrino', meaning the little neutron. Pauli also suggested correctly that the neutrino carried spin ½ to satisfy angular momentum conservation.

Pauli discussed his suggestion in many lectures given during 1931-33, which was closely followed by Fermi. Fermi also discussed this idea in depth with Pauli. This led him to publish his theory of weak interaction in 1934 [3]. This was based in analogy with Dirac's theory of electromagnetic (EM) interaction. But this was an effective theory, based on contact interaction, as illustrated in fig. 2. We know now that this contact interaction represents the modern gauge theory of weak interaction in the low energy limit, $Q^2 \ll m_W^2$, where

$$G_F = \frac{\sqrt{2}}{8}\left(\frac{g^2}{m_W^2}\right) \simeq 10^{-5} GeV^{-2} \tag{5}$$

is called the Fermi coupling. It represents the contact amplitude for weak decay processes like

$$\mu^- \to \nu_\mu e^- \bar{\nu}_e \ \& \ d \to u e^- \bar{\nu}_e, \tag{6}$$

where we have distinguished the neutrinos having charged current interaction with electron and muon by the corresponding subscripts in anticipation of the next section.

The second process describes the decay of 'down' to 'up' quark, which underlies neutron beta decay

$$n(udd) \to p(uud)e^-\bar{v}_e, \tag{7}$$

which in turn underlies all other nuclear beta decays of eq. (4).

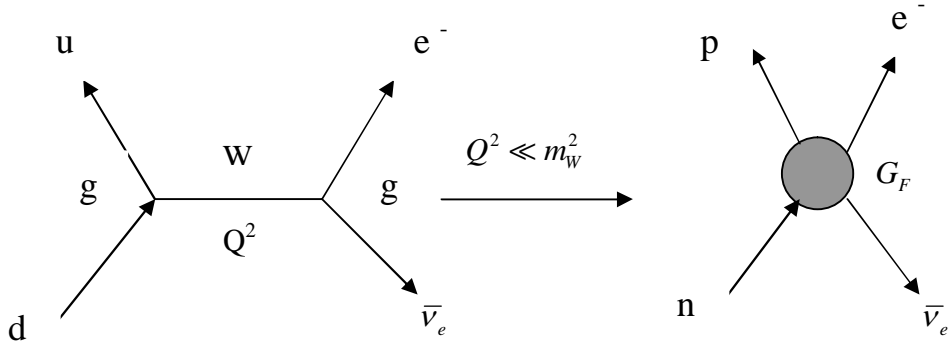

Fig. 2. Fermi's theory of weak interaction ( Four fermion contact interaction) as the low energy effective theory of charged current gauge interaction, mediated by the exchange of the massive W boson, with gauge coupling g to quarks and leptons.

One can easily compute the differential and total decay rates of neutron (eq. 7) or any radioactive nucleus (eq. 4) by using Fermi's golden rule – i.e. squaring the decay amplitude of fig.2 and integrating over the phase space of final state particles. Thus

$$d\Gamma \simeq G_F^2 \frac{d^3 p_e}{(2\pi)^3} \frac{d^3 p_v}{(2\pi)^3} (2\pi) \delta(Q - E_e - E_v) |M_W|^2,$$

$$d\Gamma/dE_e \simeq G_F^2 \frac{(4\pi)^2}{(2\pi)^5} E_e p_e \underbrace{\sqrt{(Q-E_e)^2 - m_v^2}}_{p_v} \underbrace{(Q-E_e)}_{Ev} |M_W|^2, \tag{8}$$

where $|M_W|^2 \sim 6$ is the nuclear matrix element, connecting the nucleons to the underlying quarks [1]. In the last step we have $(4\pi)^2$ from the angular integrations of $p_e$ and $p_v$, while we have used the δ-function to integrate over $E_v$. Taking $m_v = 0$, we find

$$d\Gamma/dE_e \propto p_e E_e (Q-E_e)^2 \propto \sqrt{E_e^2 - m_e^2}.E_e(Q-E_e)^2. \tag{9}$$

We can easily compute this and see that it gives a very good approximation to the shape of the electron energy spectrum for neutron or any other nuclear beta decay. The small remaining difference comes from the EM interaction of the electron (positron) with the recoiling nucleus. Finally, we can compute the total neutron decay width from (8), i.e.

$$\Gamma_n \simeq G_F^2 \int \frac{d^3 p_e}{(2\pi)^3} \frac{d^3 p_\nu}{(2\pi)^3} (2\pi) \delta\left(m_n - m_p - E_e - E_\nu\right) |M_W|^2$$
$$\simeq G_F^2 \frac{(4\pi)^2}{(2\pi)^5} \int_{m_e}^{m_n - m_p} dE_e E_e p_e \left(m_n - m_p - E_e\right)^2 |M_W|^2. \tag{10}$$

With $G_F \sim 10^{-5} GeV^{-2}$ & $|M_W|^2 \sim 6$, one finds $\Gamma_n = \hbar/\tau_n \approx 10^{-28}$ GeV, corresponding to neutron decay lifetime $\tau_n \approx 887$ s.

## 2. Neutrino scattering and detection

The Fermi theory of fig.2 describes not only neutron beta decay but also the neutrino scattering processes

$$\nu_e n \to p e^- \ \& \ \bar{\nu}_e p \to n e^+, \tag{11}$$

since in quantum field theory the absorption of a particle is equivalent to the emission of its antiparticle. These are the reactions, responsible for neutrino detection. We can easily compute the corresponding cross-section by squaring the amplitude and integrating over the 2-body phase space, i.e.

$$\sigma \simeq G_F^2 \int \frac{d^3 p_e}{(2\pi)^3} (2\pi) \delta\left(m_p + E_\nu - m_n - E_e\right) |M_W|^2$$
$$\simeq \frac{G_F^2}{\pi} p_e E_e |M_W|^2. \tag{12}$$

For incident neutrino energy significantly larger than $m_e$ we get

$$\sigma \sim G_F^2 E_\nu^2 |M_W|^2 / \pi \sim 10^{-44} cm^2 (for E_\nu \sim 1 MeV). \tag{13}$$

The above equation shows how small is the neutrino scattering cross-section. The scattering rate is given by the product of this cross-section with the incident neutrino flux times the number of target nucleons. Thus for rock, with density $\rho \sim 6$ gm/cc $\sim 10^{24}$ nucleons/cc, we see that a single neutrino in the MeV energy range will pass through $10^{20}$ cm = $10^{15}$ km of rock before any interaction. This is a hundred billion times the diameter of earth ($\approx 10^4$ km)! This shows how weak is the neutrino interaction and how hard it is to detect them.

Note that we now understand the reason for the weakness of neutrino interaction to be its short range – i.e. the heavy mass of the exchanged W boson. The range is $\hbar c/m_W \sim 0.002$ fm for the observed W boson mass of 80 GeV; while the position uncertainty of a 1 MeV neutrino is $\hbar c/1$ MeV $\sim 200$ fm. So the scattering amplitude is suppressed by the overlap fraction of the two areas $\sim (1 \text{ MeV}/M_W)^2 \sim 10^{-10}$. And the corresponding cross-section is suppressed by a factor of $10^{-20}$ with respect to the long range EM interaction. The electron has a cross-section of $\sim 10^{-24}$ cm$^2$, and gets absorbed by a few cm thick matter, in contrast with the $10^{20}$ cm for neutrino. Indeed Pauli had commented in the early 1930s that he has invented such a particle, which will never be detected experimentally.

However, neutrino was detected within Pauli's life time – i.e. in 1956 by Cowan and Reines [4]. This was made possible by the advent of nuclear reactors, which provided a rich flux of (anti)neutrinos $\sim 10^{13}$/cm$^2$/s. Their experiment consisted of two tanks, each containing 200 liters of water with dissolved Cadmium Chloride, and sandwiched between

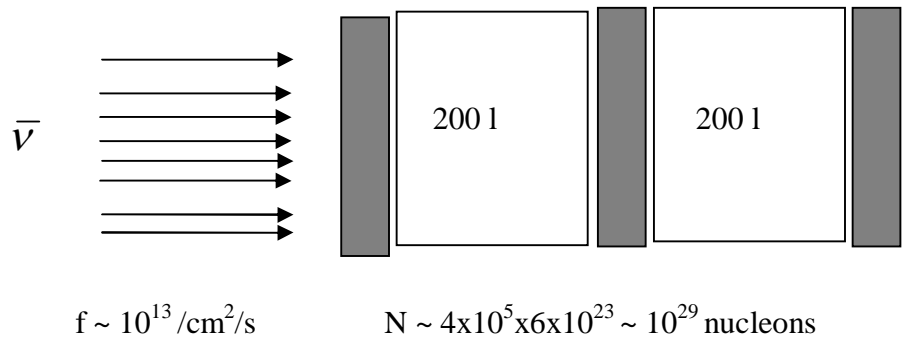

Fig.3. Schematic diagram of the neutrino discovery experiment of Cowan and Reines, showing two water tanks sandwiched between three scintillation detectors.

three scintillation detectors (Fig.3). One sees from eq. (13) that the expected interaction rate of the reactor neutrinos is

$R \sim fN\sigma \sim 10^{-2}/s,$  (14)

i.e. about 1 per 2 minutes. The proton in the water provided target for the interaction $\bar{\nu}_e p \rightarrow e^+ n$. The gamma rays from the annihilation of the produced $e^+$ with $e^-$ provided the scintillation signal. This was followed closely (within a few microseconds) by gamma ray from neutron absorption in cadmium, $n^{48}Cd \rightarrow \gamma^{49}Cd$. The observed double gamma ray signal was further confirmed by its correlation with the reactor being in operation. ( The Nobel prize came to Reines forty years later in 1996, by which time Cowan had died.)

This was followed soon by another Nobel prize winning neutrino experiment – i.e. the discovery of a second species (or flavour) of neutrino $\nu_\mu$ [5]. It is related to the muon by the charged current interaction of the type shown in fig.2. It comes from muon decay process shown in eq. (6) as well as from pion decay

$$\pi^\pm \to \mu^\pm \nu_\mu (\bar{\nu}_\mu). \qquad (15)$$

In fact this is the dominant decay mode of charged pions, as we shall see later. The high-energy neutrinos coming from this decay were bombarded on a nucleon target at the Brookhaven laboratory experiment [5], which detected the muons produced via

$$\nu_\mu n \to \mu^- p (\bar{\nu}_\mu p \to \mu^+ n). \qquad (16)$$

That the produced particle was $\mu^\pm$ instead of $e^\pm$ demonstrated the existence of a second species of neutrino. This was closely followed by the first detection of $\nu_\mu$ in a cosmic ray experiment at the Kolar gold mine [6], which was in fact the first detection of atmospheric neutrino. Finally the third neutrino species (flavour) $\nu_\tau$ was discovered in 2000 at Fermilab by observing the $\tau$ leptons produced via

$$\nu_\tau n \to \tau^- p (\bar{\nu}_\tau p \to \tau^+ n) \qquad (17)$$

in a nuclear emulsion experiment [7].

## 3. Neutrino chirality and mass:

The abovementioned quarks, charged leptons and neutrinos are all spin 1/2 particles, called matter fermions. So they can occur in right- and left-handed chirality (or helicity) states, corresponding to spin alignment along and opposite to the direction of momentum. As per the Standard Model of particle physics the three basic interactions between these particles – strong, electromagnetic and weak – are all gauge interactions. Their strengths are determined by the respective gauge charges – i.e colour charge, electric charge and weak isospin. The strong and the EM interactions are invariant under charge conjugation C and parity transformation P (i.e. space reflection). However, it was discovered in the 1950s that the weak interaction breaks C and P maximally, while preserving invariance under the combined CP transformation. This means that only the left-handed quarks and leptons (and the corresponding right-handed antiparticles) carry weak isospin (I=1/2) and take part in weak interaction. Thus we have three isospin doublets of left-handed leptons and quarks

$$\begin{pmatrix} \nu_e \\ e^- \end{pmatrix}_L \begin{pmatrix} \nu_\mu \\ \mu^- \end{pmatrix}_L \begin{pmatrix} \nu_\tau \\ \tau^- \end{pmatrix}_L \& \begin{pmatrix} u \\ d \end{pmatrix}_L \begin{pmatrix} c \\ s \end{pmatrix}_L \begin{pmatrix} t \\ b \end{pmatrix}_L \qquad (18)$$

and similarly for their right-handed antiparticles. On the other hand the right-handed quarks and charged leptons occur as isospin singlets (I=0) along with their left-handed antiparticles; and take part only in strong and EM interactions. Since the neutrinos have only weak interaction there is no need for right-handed neutrinos (or left-handed antineutrinos); and there are none in the Standard Model.

Now a fermion mass term in the Lagrangian corresponds to the combination

$$m_f \bar{\psi}_{R(L)} \psi_{L(R)} = \bar{f}_{L(R)} \xrightarrow{\Leftarrow(\Rightarrow)} \xleftarrow{\Rightarrow(\Leftarrow)} f_{L(R)}, \tag{19}$$

where $\psi_L \& \bar{\psi}_L$ are wavefunctions for left-handed fermion and its antiparticle – i.e. right-handed antifermion. (In the more rigorous language of quantum field theory $\psi_L \& \bar{\psi}_R$ are field operators for absorbing left-handed fermion and antifermion or creating right-handed antifermion and fermion respectively). The long and short arrows on the right hand side illustrate the momenta and spins of the antifermion and fermion pair in their centre of mass frame. This illustration shows that only these two combinations are allowed by angular momentum conservation, while LR and RL terms are disallowed because they carry nonzero total angular momentum, which will violate the rotational invariance of the Lagrangian. In other words, the mass term represents the absorption or creation of a fermion-antifermion pair of same chirality LL or RR (or transformation of a left-handed fermion into a right-handed one, i.e. chiral symmetry breaking).

Note that in the Standard Model a left-handed (or right-handed) fermion-antifermion pair carries total isospin 1/2; and hence breaks gauge invariance of the Lagrangian. So even the quarks and charged leptons cannot have bare mass, represented by the above mass term in the Lagrangian. Instead they get mass via their Yukawa coupling to the isospin doublet of Higgs boson, which acquires a vacuum expectation value by spontaneous breaking of the isospin gauge symmetry, i.e.

$$y\bar{\psi}_R \psi_L h \xrightarrow{sp.symm.br} \underbrace{y\langle h \rangle}_{m} \bar{\psi}_R \psi_L. \tag{20}$$

In other words their mass comes from their Yukawa interaction with the constant Higgs field <h>, present in the vacuum. This is called the Dirac mass of quarks and charged leptons, which is roughly in the range of $\sim 10^{\pm 2}$ GeV.

It is worth making a small detour on Spontaneous Symmetry Breaking, which means a symmetry is preserved by the Lagrangian but broken by the ground state. Here the Isospin symmetry of the Lagrangian is broken by the ground state (vacuum) due to the presence of the constant Higgs field, carrying nonzero isospin (I=1/2). The simplest example of SSB is a ferromagnet, where the rotational symmetry of the electro-magnetic interaction Lagrangian is broken by the ground state, as its atomic spins get aligned with one another in a particular (north-south) direction. We have a similar situation in Higgs mechanism, except that the rotational symmetry is in Isospin space instead of the ordinary space. Experimental evidence for this is expected to come from the discovery of Higgs boson at the large Hadron Collider. Incidentally, there is a spontaneous breaking of chiral symmetry in the strong interaction of quarks, which accounts for the bulk of the nucleon mass. This was the subject of the 2008 Nobel prize in physics to Nambu.

However, since there is no right-handed neutrino or left-handed antineutrino in the Standard Model, there cannot be any neutrino mass as per eq. (19). But one can extend the Standard Model by introducing an isospin singlet $v_R(\bar{v}_L)$ like the other fermions, since it does not violate any basic symmetry of this model. Unlike the other fermions, however, the singlet $v_R(\bar{v}_L)$ has a unique property, i.e. it does not carry any gauge charge. Thus the particle and antiparticle are indistinguishable except for their opposite chiralities. Although they carry opposite lepton numbers, it is not a gauge quantum number and hence not required to be conserved. Hence such particles can get a new kind of mass called Majorana mass, i.e.

$$M\psi_{vR}\psi_{vR,} \tag{21}$$

representing absorption of two right-handed netrinos. Moreover this mass can be arbitrarily large, since it does not break any gauge symmetry of the Standard Model, unlike the Dirac mass (20). Note however that it breaks lepton number conservation by two units (ΔL=2). Apart from this the neutrino can also have a Dirac mass (20) now like the other fermions. Thus one has a 2x2 mass matrix in the basis of doublet and singlet neutrinos

$$\begin{pmatrix} 0 & m \\ m & M \end{pmatrix} \rightarrow \begin{pmatrix} m^2/M & 0 \\ 0 & M \end{pmatrix}, \tag{22}$$

which induces a tiny mass for the standard doublet neutrino on diagonalisation. This is called the see-saw model, since the larger the Majorana mass the smaller will be the standard doublet neutrino mass induced by it. Note that in contrast the other fermions have no Majorana mass, so that on diagonalisation both the singlet and the doublet components have the same mass m.

There is a great deal of current interest in pursuing this model for two reasons. (I) In an important extension of the Standard Model, called grand unified theory (GUT), the lepton number L can be a gauge charge. In that case the Majorana mass M will represent the GUT symmetry breaking scale, just like the Dirac mass represents the spontaneous symmetry breaking scale of the Standard Model. (II) The lepton number violation, associated with a large Majorana mass scale, can generate a lepton asymmetry in the early universe, which will be able to explain the present baryon asymmetry of the universe. In either case one can get an indirect indication of physics at an inaccessibly high mass scale by probing the tiny but nonzero doublet neutrino mass induced by it. We shall see below that there are indeed strong experimental evidences of such a tiny neutrino mass in the sub-eV range, which is indicative of new physics in the ultrahigh mass scale of $M \geq 10^{10} GeV$.

## 4. Neutrino mixing and oscillation

If the neutrinos have non-zero masses then there is no reason for the three neutrino interaction (or flavour) eigenstates of eq. (18) to coincide with the three mass eigenstates. In general there will be mixing between them. To a good approximation it is enough to consider mixing between two neutrino states, e.g.

$$\begin{pmatrix} \nu_e \\ \nu_\mu \end{pmatrix} = \begin{pmatrix} \cos\theta & \sin\theta \\ -\sin\theta & \cos\theta \end{pmatrix} \begin{pmatrix} \nu_1 \\ \nu_2 \end{pmatrix}, \qquad (23)$$

where $\nu_1$ and $\nu_2$ are mass eigenstates with eigenvalues $m_1$ and $m_2$. Note that coherent mixing between the mass eigenstates is a quantum mechanical phenomenon. The mixed state

$$\nu_e = \nu_1 \cos\theta + \nu_2 \sin\theta \qquad (24)$$

breaks energy conservation. This is most easily seen in the rest frame of the lighter mass eigenstate, corresponding to $m_1$ say. Then in this frame the $\nu_2$ component has a higher energy, $m_2$ (plus any kinetic energy). Such a coherent admixture of unequal mass eigenstates is possible in quantum mechanics, where the energy nonconservation problem is taken care of by the uncertainty principle. This leads to the phenomenon of neutrino oscillation, as pointed out by Pontecorvo [8].

Consider a $\nu_e$ state produced via a nuclear beta decay. Its $\nu_1$ and $\nu_2$ components will travel with different velocities as they have different masses. Therfore their relative sizes will change with distance, which means transformation of $\nu_e$ into $\nu_\mu$. Of course neutrinos of definite mass m and momentum p do not travel as point particles due to the uncertainty principle. Instead they travel as a plain monochromatic wave, represented by the wave function

$$\psi = e^{-i(Et-pl)}. \qquad (25)$$

Moreover, since E (~ MeV) is much larger than the neutrino mass (< eV), they are extreme relativistic particles, i.e.

$$t \simeq l \ \& \ E \simeq p + m^2/2p \simeq p + m^2/2E \qquad (26)$$

in natural units. Substituting (26) in (25) we see that the neutrino mass eigenstate propagates with a phase of $\exp(-im^2 l/2E)$. Thus the wave function of the produced $\nu_e$ of eq. (24) after traveling a distance $l$ becomes

$$\nu_e \to \nu_1 \cos\theta\, e^{-im_1^2 l/2E} + \nu_2 \sin\theta\, e^{-im_2^2 l/2E}. \qquad (27)$$

Now decomposing the $\nu_1$ and $\nu_2$ components back into $\nu_e$ and $\nu_\mu$ via eq. (23) we see that the coefficient of the $\nu_\mu$ term does not cancel out. Indeed the modulus square of this coefficient gives the probability of $\nu_e$ oscillation into $\nu_\mu$

$$P_{e\mu}(l) = \left|\cos\theta \sin\theta \left(-e^{-im_1^2 l/2E} + e^{-im_2^2 l/2E}\right)\right|^2$$
$$= \sin^2 2\theta \sin^2\left(\Delta m^2 l/4E\right), \tag{28}$$

where $\Delta m^2 = m_2^2 - m_1^2$. One can convert this from natural units to more convenient units of $\Delta m^2 (eV^2)$, $l$ (meter) and E (MeV), giving

$$P_{e\mu}(l) = \sin^2 2\theta \sin^2\left(1.3\Delta m^2 l/E\right). \tag{29}$$

Note that the first factor gives the amplitude and the second factor the phase of neutrino oscillation. From this phase one gets the wavelength of neutrino oscillation

$$\lambda = (\pi/1.3)(E/\Delta m^2) \simeq 2.4 E/\Delta m^2. \tag{30}$$

Thus for large mixing $(\sin^2 2\theta \sim 1)$ one expects the following pattern of neutrino oscillation probability from eqs. (29,30), where the factor of 1/2 in the last case comes from averaging over the phase factor.

| $l$ | $\ll\lambda$ | $\sim\lambda/2$ | $\gg\lambda$ |
|---|---|---|---|
| $P_{e\mu}$ | 0 | $\sin^2 2\theta \sim 1$ | $(1/2)\sin^2 2\theta \sim 1/2$ |

(31)

Note that the corresponding survival probability is given by the remainder, i.e.

$$P_{ee} \equiv P_{\nu_e \to \nu_e} = 1 - P_{e\mu}. \tag{32}$$

For solar or reactor neutrino experiment the source of $\nu_e$ is nuclear reaction, and hence the energy E ~ MeV. The distance between the reactor (source) and the detector for a long baseline experiment like KamLAND is a few hundred km ($l \sim 10^5$ m), while for solar neutrino experiment $l \sim 10^{11}$ m. Thus one sees from eqs. (30-32) that KamLAND and solar neutrino experiments can probe neutrino mass down to $\Delta m^2 \sim 10^{-5}$ and $10^{-11}$ eV$^2$ respectively, which is far beyond the reach of any other method of mass measurement.

It should be noted here that the typical energy for accelerator and atmospheric neutrinos is E ~ GeV. Thus one can use the same eqs. (29,30) by measuring distance in km. For a long baseline accelerator neutrino experiment like MINOS the distance between the source and the detector is $l \sim 10^3$ km, while for atmospheric neutrinos traversing the earth it is given by the diameter of the earth, $l \sim 10^4$ km. Thus these experiments can probe neutrino masses down to $\Delta m^2 \sim 10^{-3}$ and $10^{-4}$ eV$^2$ respectively. We shall see below that the atmospheric/accelerator neutrino experiments as well as the solar/reactor neutrino experiments give unambiguous evidence of neutrino masses in the sub-eV range.

## 5. Atmospheric neutrino oscillation

The high energy cosmic rays collide with the nuclei of the earth's atmosphere producing π and K mesons. The decay of these mesons produce $\mu^{\pm} + \nu_{\mu}(\bar{\nu}_{\mu})$ as per eq. (15). This dominates over the $e^{\pm} + \nu_e(\bar{\nu}_e)$ decay mode, although the latter is kinematically favoured. The reason is that in the rest frame of the π (K) meson the $e^{\pm}$ is an extreme relativistic particle, so that the helicity of the daughter particles are the same as their chirality. Since only left-handed leptons and right-handed antileptons take part in weak interaction, we get the following final state particles along with their helicities in the π rest frame:

$$e_R^+ \xleftarrow{\Leftarrow} \pi^+ \xrightarrow{\Leftarrow} \nu_{eL}. \tag{33}$$

Note that the net spin projection of the $e^+\nu_e$ pair along the direction of motion is S=1. Since π meson has no spin and the orbital angular momentum has no projection along the direction of motion, this decay is disallowed by angular momentum conservation. The corresponding $\mu^+\nu_\mu$ decay is allowed because μ is non-relativistic in the π rest frame, and hence its chirality is not the same as its helicity. Thus we expect to see the decays

$$\pi^{\pm} \to \mu^{\pm} + \nu_{\mu}(\bar{\nu}_{\mu}), \mu^{\pm} \to \bar{\nu}_{\mu} e^+ \nu_e (\nu_{\mu} e^- \bar{\nu}_e), \tag{34}$$

i.e. $\nu_e(\bar{\nu}_e) : \nu_\mu(\bar{\nu}_\mu)$ in the ratio 1:2. ( At higher energy this ratio may be < 1/2 as the μ may not decay in the atmosphere). The main atmospheric neutrino experiment is the SuperKamioka neutrino detection experiment (SK), located in the Kamioka mines in Japan. It consists of a 50 kiloton water Cerenkov detector, surrounded by thousands of photomultiplier tubes to catch the Cerenkov radiation. The electron (muon) neutrinos produce electrons (muons) by charged current interactions (11,16) in this water, which produce Cerenkov radiation. Since the Cerenkov ring produced by an energetic electron is more diffused than the ring produced by a muon, the experimenters can distinguish them to a very good accuracy, although they cannot determine the lepton charge.

From the measured rate of electron and muon production they estimated the $\nu_e(\bar{\nu}_e)$ and $\nu_\mu(\bar{\nu}_\mu)$ fluxes. While the former was in agreement with the expectation from the known cosmic ray fluxes, the latter showed a clear deficit, indicating $\nu_\mu \to \nu_\tau$ oscillation. Fig.4 shows the zenith angle distribution of the observed electron and muon like events along with the corresponding theoretical predictions with and without neutrino oscillation [9]. The observed e-like events agree with the no oscillation prediction. However, the observed μ-like event rate shows a clear deficit for both sub-GeV and multi-GeV neutrino energies. Note that the sub-GeV neutrinos have a relatively small oscillation wavelength λ from eq. (30), which accounts for the deficit being seen for neutrinos at all zenith angles. In contrast the multi-GeV neutrinos have a large oscillation wavelength, which is responsible for the effect being seen only for upward going (earth traversing) neutrinos.

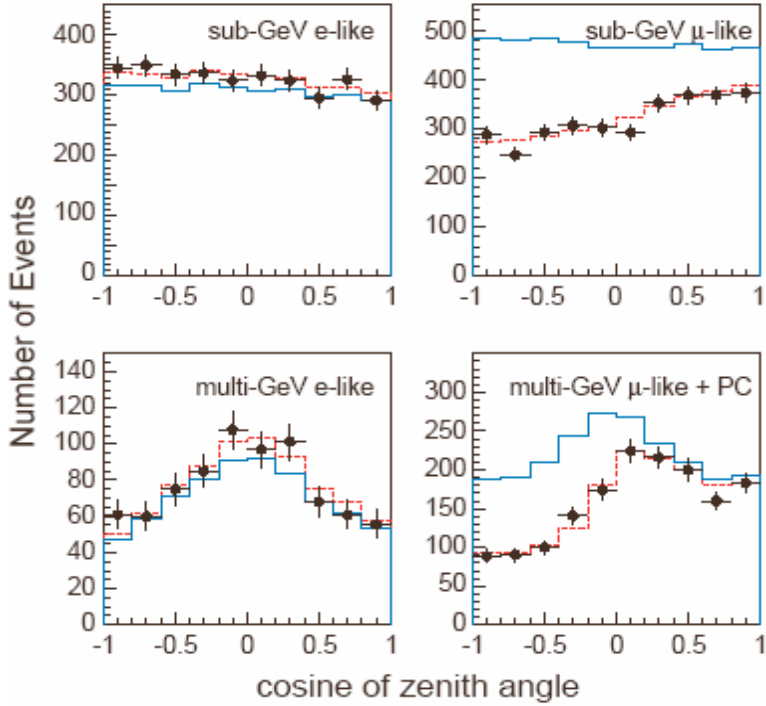

Fig.4. The zenith angle distribution of the SK electron and muon like events along with the theoretical expectations without (blue continuous lines) and with (red dashed lines) neutrino oscillation [9].

For such neutrinos the distance traveled is simply related to the zenith angle via

$$l \simeq -D\cos\theta, \qquad (35)$$

where D ~ 13,000 km is the diameter of earth. Thus one can measure both the energy and the distance traveled by the neutrino, and look for the oscillatory pattern of the predicted survival probability of eqs. (29) and (32) as a function of the ratio $l/E$. This is shown in fig.5 [10].

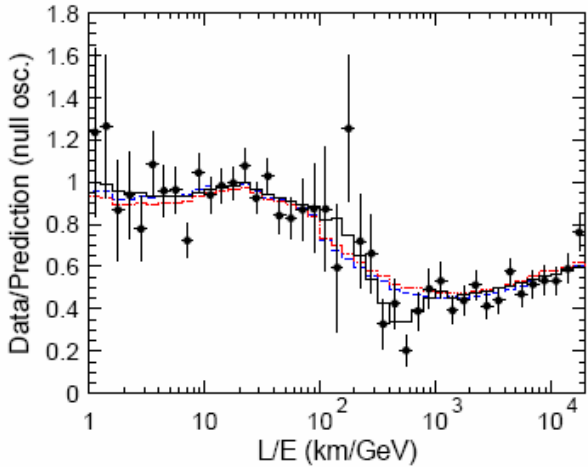

Fig.5. The SK muon like event rates relative to the theoretical prediction without oscillation (i.e. the $\nu_\mu$ survival probability) is shown as a function of the ratio $l/E$ along with the best $\nu_\mu \to \nu_\tau$ oscillation fit (black solid line) [10]. Some alternative model fits in terms of neutrino decay and decoherence are also shown for comparison.

One can clearly see the survival minimum (oscillation maximum) at

$$l/E = \lambda/2E = 1.2/\Delta m^2 \simeq 500\, km/GeV. \tag{36}$$

This corresponds to $\Delta m^2 \simeq 2.4 \times 10^{-3} eV^2$, while the large oscillation amplitude implies $\sin^2 2\theta \simeq 1, i.e. \theta \simeq \pi/4$.

This result has been confirmed now by two long baseline accelerator neutrino experiments in Japan (K2K) [11] and USA (MINOS) [12]. The K2K experiment uses the neutrino beam from the KEK accelerator and the SK detector, separated by a distance of 250 km. MINOS uses the neutrino beam from Fermilab accelerator and an iron detector, separated by a distance of 730 km. Fig.6 shows the consistency of the atmospheric neutrino oscillation parameter from all the three experiments [12].

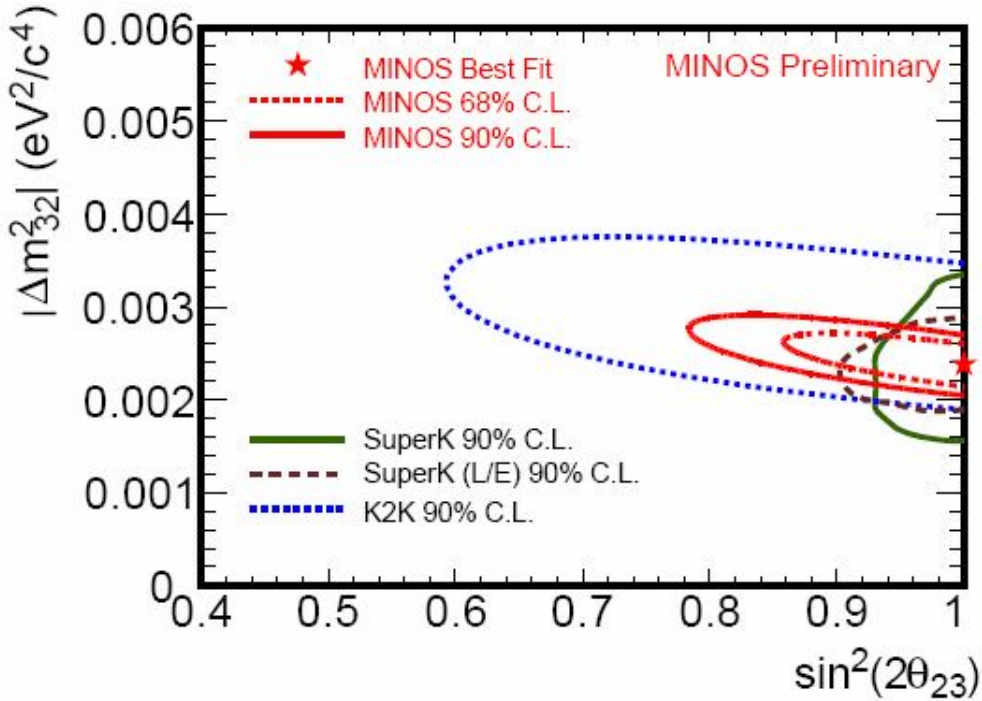

Fig.6. Confirmation of the atmospheric neutrino oscillation parameters of the SK experiment by the MINOS and the K2K long baseline experiments [12].

The best values of the atmospheric neutrino mass and mixing parameters are

$$\left|\Delta m_{atm}^2\right| = \left|\Delta m_{32}^2\right| \simeq 2.4 \times 10^{-3} eV^2, \sin^2 2\theta_{atm} = \sin^2 2\theta_{23} \simeq 1. \tag{37}$$

## 6. Solar neutrino oscillation

The main source of solar neutrino as well as solar energy are the three pp chains of nuclear reactions, which convert protons into $^4$He (alpha particle). These reactions take place in the solar core, as first suggested by Bethe in 1939. They are shown in fig.7. While most of this conversion takes place via the shortest chain (pp-I), a small fraction (15%) follows a detour via $^7$Be (pp-II); and a tiny fraction (0.1%) of the latter follows a still longer detour via $^8$B (pp-III). The resulting neutrinos are (I) the low energy pp neutrino, (II) the intermediate energy Be neutrino and (III) the relatively high energy B neutrino, with decreasing order of flux.

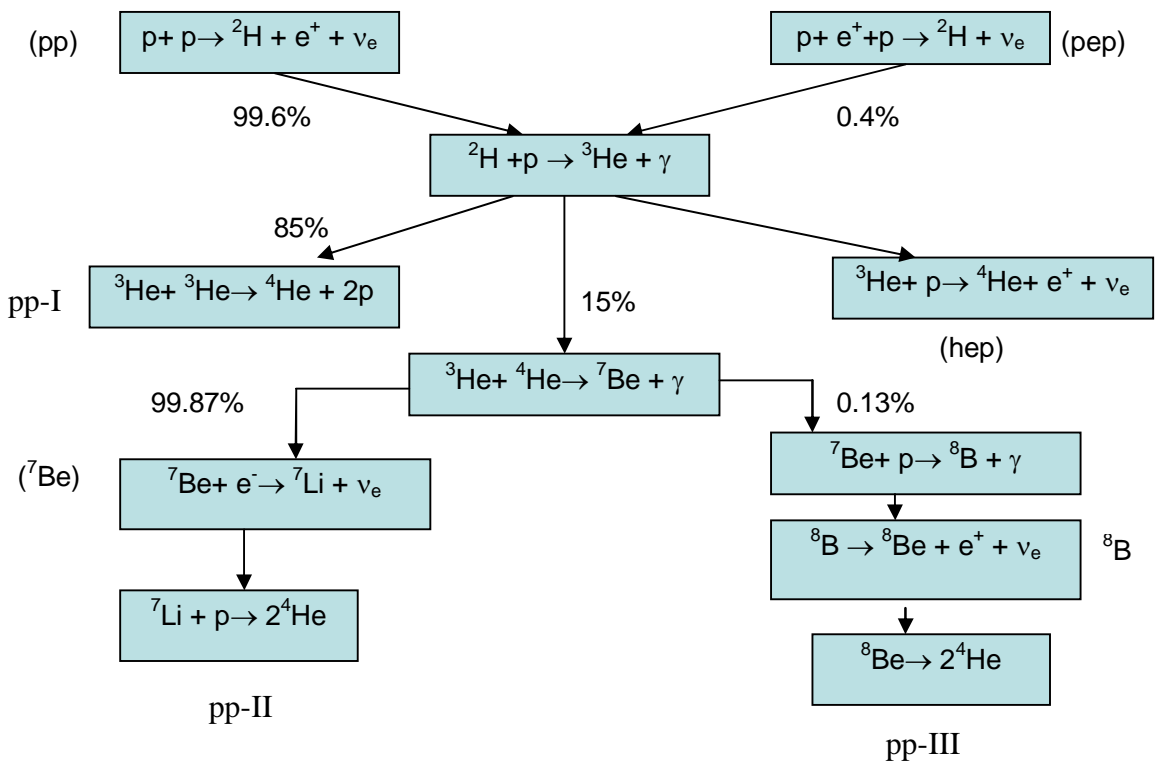

Fig.7. The three pp chains of nuclear reactions, which constitute the most important source of solar neutrinos as well as solar energy.

The standard solar model prediction for these neutrino fluxes are shown in fig.8 [13]. It also shows the neutrino energy ranges covered by the different solar neutrino experiments, described below.

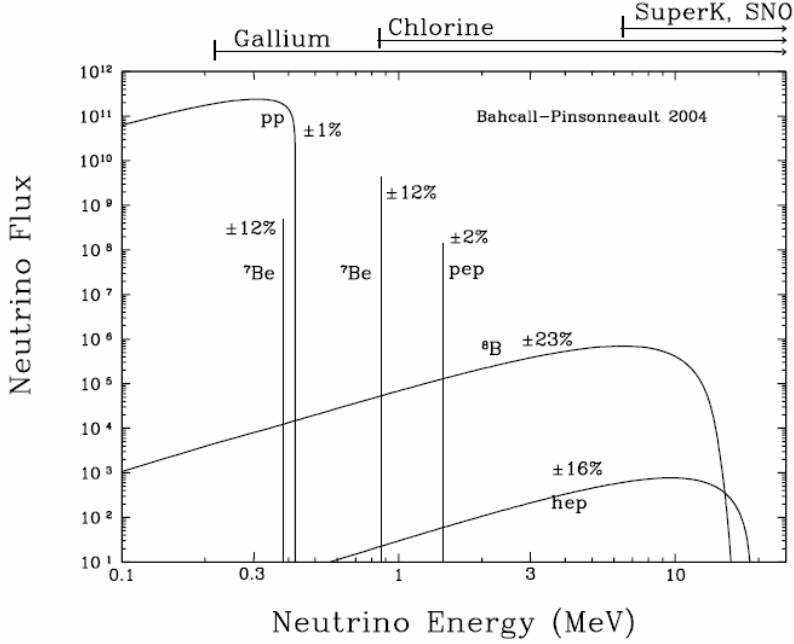

Fig.8. The standard solar model (SSM) prediction for the solar neutrino fluxes shown along with the energy ranges of the solar neutrino experiments [13].

The Gallium [14] and Chlorine [15] experiments are radio chemical experiments, based on the charged current reactions

$$\nu_e + {}^{71}Ga \rightarrow e^- + {}^{71}Ge, \nu_e + {}^{37}Cl \rightarrow e^- + {}^{37}Ar. \qquad (38)$$

The produced $^{71}$Ge and $^{37}$Ar are periodically extracted and measured by radiochemical method, from which the incident neutrino fluxes are estimated. The value of this observed flux R relative to the SSM prediction gives the $\nu_e$ survival probability $P_{ee}$. The SK water Cerenkov experiment [16] is a real time solar neutrino experiment, which is based on elastic scattering of $\nu_e$ on electron. It is dominated by the charged current interaction $\nu_e e^- \rightarrow e^- \nu_e$, but it also has a limited sensitivity to the neutral current interaction $\nu_{e,\mu} e^- \rightarrow \nu_{e,\mu} e^-$. Thus

$$R_{el} = P_{ee} + \frac{\sigma^{NC}}{\sigma^{NC} + \sigma^{CC}}(1 - P_{ee}) \simeq P_{ee} + \frac{1}{6}(1 - P_{ee}). \qquad (39)$$

This experiment can also measure the energy and direction of the incident $\nu_e$ from those of the outgoing electron. The Sudbury Neutrino Observatory (SNO) is also a Cerenkov experiment like SK, but with a heavy water target [17,18]. Since deuteron has a low binding energy of only 2.2 MeV, it can detect both charged and neutral current interactions via

$$\nu_e + d \xrightarrow{CC} e^- + p + p,$$
$$\nu_{e,\mu} + d \xrightarrow{NC} \nu_{e,\mu} + p + n. \tag{40}$$

In the phase-I of this experiment the NC events were detected via neutron capture on deuteron, n + d → t + γ [17]. In the phase-II salt was added to the heavy water target to enhance the NC detection efficiency via neutron capture on Chlorine, n + $^{35}$Cl→$^{36}$Cl + γ [18]. In both cases the γ is detected via the Cerenkov radiation produced by Compton scattering. Consequently there is some ambiguity in separating the charged and neutral current interactions; but it is much less for phase-II. In the last phase of this experiment, $^3$He gas filled counters were inserted into the heavy water target to detect the NC events via counting the protons from n + $^3$He → p + t.

Table 1 shows the energy threshold of the above four experiments along with the compositions of the corresponding solar neutrino spectra. It also shows the corresponding survival probability $P_{ee}$ measured by the rates of the charged current reaction relative to the SSM prediction. For the SK experiment the survival probability calculated from $R_{el}$ via eq. (39) is shown in parenthesis. It may be noted here that both the Chlorine and the Kamioka experiments have got Nobel prize for detecting solar neutrino signal. But while the observed rates are clearly nonzero they are also clearly much below the SSM prediction. It is this latter aspect that concerns us here; and the Nobel prize for this is yet to come. We see from Table 1 that the survival probability of $\nu_e$ is slightly above 1/2 for the low energy solar neutrino, falling to 1/3 at higher energy. To understand its magnitude and energy dependence we have to consider the effect of solar matter on neutrino oscillation.

**Table 1.** The $\nu_e$ survival probability $P_{ee}$ measured by the CC event rate R of various solar neutrino experiments relative to the SSM prediction. For SK the Pee obtained after the NC correction is shown in parenthesis.

| Experiment | Gallium | Chlorine | SK | SNO-I |
|---|---|---|---|---|
| R | 0.55 ± 0.03 | 0.33 ± 0.03 | 0.465 ± 0.015 (0.36 ± 0.015) | 0.35 ± 0.03 |
| $E_{th}$ (MeV) | 0.2 | 0.8 | 5 | 5 |
| Composition | pp (55%), Be (25%), B (10%) | B (75%), Be (15%) | B (100%) | B (100%) |

**Matter Enhancement (Resonant Conversion):**

We know that the photon gets an induced mass while propagating in glass or water, which is responsible for its refractive index. It comes from the interaction of photon with the material of the medium. Likewise the neutrino gets an induced mass from its interaction with solar matter, which has a profound effect on its oscillation. It is known as MSW effect after its authors [19]. It arises from the charged current interaction of $\nu_e$ with the solar electrons, $\nu_e e^- \xrightarrow{CC} e^- \nu_e$, while the neutral current interaction has no net effect since it is common to all neutrino flavours. The CC interaction is identical to the contact four-fermion interaction shown in fig.2. The corresponding potential energy density is

$$V = \sqrt{2} G_F N_e, \tag{41}$$

where $N_e$ is the local electron density of the sun.

We have to add this potential energy term to the wave equation of the free neutrino wavefunction of eq. (25), i.e.

$$i d\psi_{1,2}/dt = (p + m_{1,2}^2/2E)\psi_{1,2}. \tag{42}$$

It is convenient for this purpose to rewrite this equation in the neutrino flavour basis using the rotation matrix (23). We get

$$\frac{id}{dt}\begin{pmatrix} \nu_e \\ \nu_\mu \end{pmatrix} = (p + M^2/2E)\begin{pmatrix} \nu_e \\ \nu_\mu \end{pmatrix}, \tag{43}$$

where

$$M^2 = \begin{pmatrix} c & s \\ -s & c \end{pmatrix}\begin{pmatrix} m_1^2 & 0 \\ 0 & m_2^2 \end{pmatrix}\begin{pmatrix} c & -s \\ s & c \end{pmatrix}, \tag{44}$$

and s, c denote sinθ, cosθ. Adding the potential energy term to the RHS of (43) is equivalent to replacing $M^2$ by an effective mass (or energy) term

$$\begin{aligned}M'^2 &= \begin{pmatrix} c & s \\ -s & c \end{pmatrix}\begin{pmatrix} m_1^2 & 0 \\ 0 & m_2^2 \end{pmatrix}\begin{pmatrix} c & -s \\ s & c \end{pmatrix} + \begin{pmatrix} 2\sqrt{2}EG_F N_e & 0 \\ 0 & 0 \end{pmatrix} \\ &= \begin{pmatrix} c^2 m_1^2 + s^2 m_2^2 + 2\sqrt{2}EG_F N_e & -sc\Delta m^2 \\ -sc\Delta m^2 & c^2 m_2^2 + s^2 m_1^2 \end{pmatrix}.\end{aligned} \tag{45}$$

So the problem reduces to finding the eigenvalues and eigenstates of this matrix.

To simplify it further let us assume for a moment that $\sin\theta \ll 1$, so that the nondiagonal elements are very small. Then we can approximately identify each diagonal element with an eigenvalue and the corresponding eigenstates $\nu_{1,2}$ with the flavour eigenstates $\nu_{e,\mu}$. Fig.9 shows the two eigenvalues as functions of the solar electron density, assuming $m_1 < m_2$. At the surface of the sun ($N_e=0$) the 1st eigenvalue ($m_1^2$) is smaller than the 2nd ($m_2^2$). But $\lambda_1$ increases steadily with $N_e$ and becomes larger than $\lambda_2$ at the solar core. The cross-over occurs at the critical density

$$N_e^C = \frac{\Delta m^2}{2\sqrt{2}G_F E}\cos 2\theta \Rightarrow M'^2_{11} = M'^2_{22}. \tag{46}$$

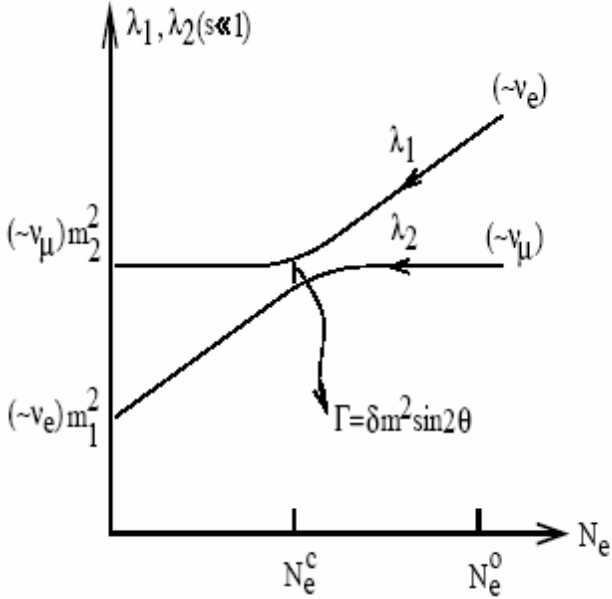

Fig.9. Schematic diagram of the two effective mass (energy) eigenvalues as functions of the solar electron density.

Note however that the two actual eigenvalues never cross. There is a minimum gap between them given by the nondiagonal element

$$\Gamma = \Delta m^2 \sin 2\theta. \tag{47}$$

It means that a $\nu_e$ produced at the solar core will come out as $\nu_2$ provided the transition probability between the two energy levels remains small.

It is easy to show that this remarkable result does not depend on the $\sin\theta \ll 1$ assumption. Consider the effective mixing angle $\theta_M$ in matter, which diagonalises the above matrix, i.e.

$$\tan 2\theta_M = \frac{2M'^2_{12}}{\left|M'^2_{22} - M'^2_{11}\right|} = \frac{\sin 2\theta}{\left|\cos 2\theta - 2\sqrt{2}G_F E N_e/\Delta m^2\right|}. \tag{48}$$

The electron density at the solar core is $N_e^0 \gg N_e^C$, so that the second term in the denominator is much larger than the first. This means $\theta_M \ll 1$ at the solar core for any vacuum mixing angle θ, so that the $\nu_e$ produced there is dominated by the $\nu_1$ component. At the critical density the denominator of eq. (48) vanishes, which corresponds to maximal mixing between the two components ($\theta_M=\pi/4$), again for any value of θ. This is why it is called matter enhanced (or resonant) conversion. The neutrino comes out from the sun as $\nu_2$ with

$$P_{ee} = \sin^2\theta, \tag{49}$$

provided the transition probability between the two levels remains small throughout the propagation. The most important region for this transition is the critical density region, where the gap between the two levels is the smallest. This transition probability is given by the Landau-Zenner formula, which was derived for quantum transition between atomic energy levels in the 1930s. It relates the transition rate T to the above gap via

$$T = e^{\frac{-\pi}{2\gamma}}, \gamma = \frac{\lambda_C (d\lambda_1/dl)_C}{\Gamma} \propto \frac{\lambda_C (dN_e/dl)_C}{N_e^C}, \tag{50}$$

where $\lambda_C$ represents the oscillation wavelength in matter in the critical density region. If the solar electron density varies so slowly that the resulting variation in the 1$^{st}$ eigenvalue over an oscillation wavelength is small compared to the gap between the two, then $\gamma \ll 1$ and the transition rate is exponentially suppressed. This is called the adiabatic condition. Thus the two conditions for the solar $\nu_e$ to emerge as $\nu_2$ are given by

$$\frac{\Delta m^2 \cos 2\theta}{2\sqrt{2} G_F N_e^0} < E < \frac{\Delta m^2 \sin^2 2\theta}{2\cos 2\theta (dN_e/dlN_e)_C}, \tag{51}$$

where the 1$^{st}$ inequality ensures $N_e^0 > N_e^C$ and the 2$^{nd}$ is the adiabatic condition.

Fig.10 shows the triangular region in the $\Delta m^2$ – $\sin^2 2\theta$ plot satisfying the above two conditions for a typical solar neutrino energy E = 1 MeV [20]. The horizontal side of the triangle follows from the 1$^{st}$ inequality, which gives a practically constant upper limit of $\Delta m^2$ in terms of the solar core electron density, since $\cos 2\theta \approx 1$. The 2$^{nd}$ inequality (adiabatic condition) gives a lower limit on $\sin^2 2\theta$, determined by the solar electron density gradiant. Moreover since this condition implies a lower limit on the product $\Delta m^2 \sin^2 2\theta$, it corresponds to a diagonal line on the log-log plot. The vertical side of the triangle is simply the physical boundary corresponding to maximal mixing. This is the so called MSW triangle.

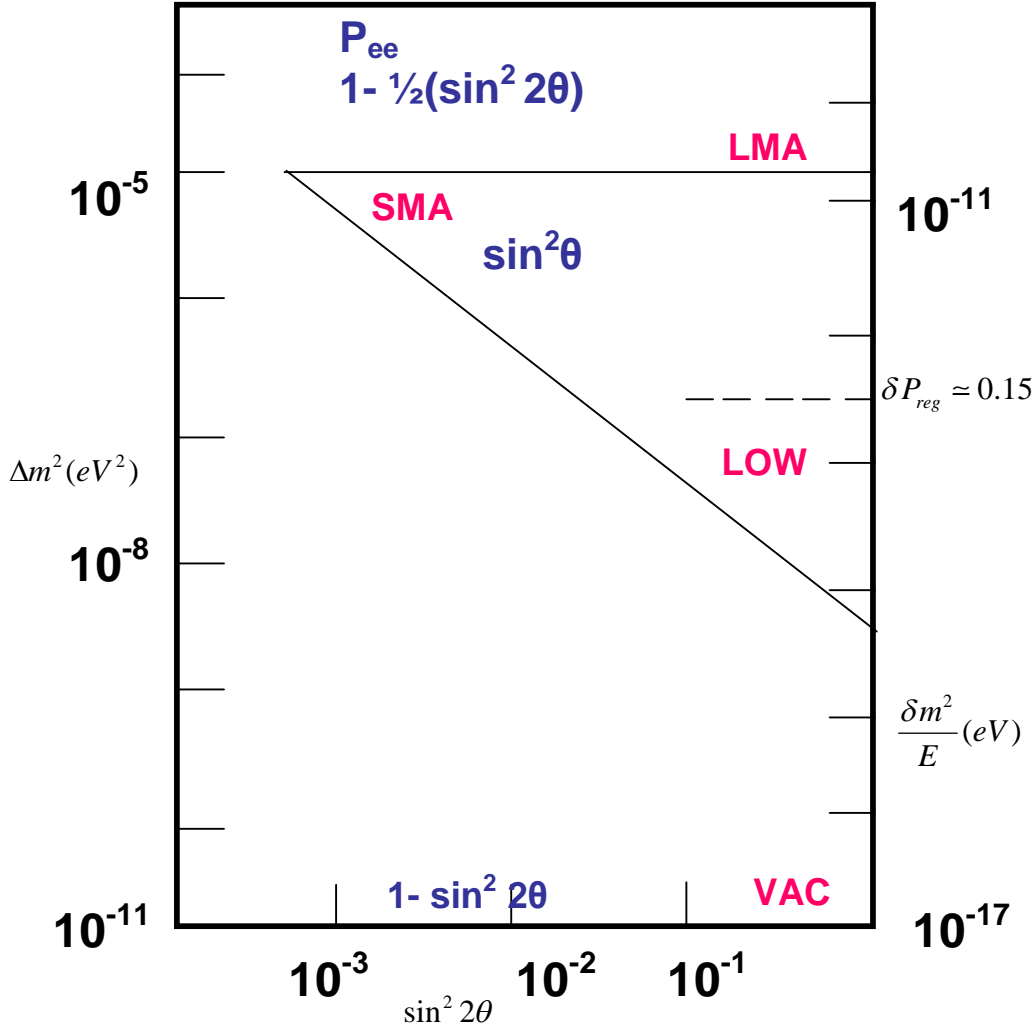

Fig10. The positions of the MSW triangle, the earth regeneration effect and the vacuum oscillation maximum are shown for E = 1 MeV along with the positions of the SMA, LMA, LOW and VAC solutions. While the former positions scale with energy the latter ones are independent of it.

The indicated survival probabilities outside the triangle follows from vacuum oscillation formulae (31,32), while that inside corresponds to eq. (49). Thus Pee is < 1/2 inside the MSW triangle and > 1/2 outside it, except for the oscillation maximum at the bottom, where the survival probability goes down to $\cos^2 2\theta$. Finally, the earth matter effect gives a small but positive $\nu_e$ regeneration probability, which means the sun shines a little brighter at night in the $\nu_e$ beam. After day-night averaging one expects a $\nu_e$ regeneration probability

$$\delta P_{reg} = \frac{\eta_E \sin^2 2\theta}{4\left(1 - 2\eta_E \cos 2\theta + \eta_E^2\right)}, \eta_E = \frac{2}{3\rho Y_e}\left(\frac{\Delta m^2/E}{10^{-13} eV}\right), \quad (52)$$

where ρ is the matter density of earth in gm/cc and Ye is the average number of electrons per nucleon. For favourable values of $\Delta m^2$ and θ, the $\delta P_{reg}$ can go up to 0.15 as indicated in the figure. It is important to note that the positions of the MSW triangle, the earth regeneration region and the vacuum oscillation maximum depend only on the ratio $\Delta m^2/E$, as one can see from the corresponding formulae. Thus their positions on the right hand scale of the figure hold at all energies.

**Four Alternative Solutions and their Experimental Resolution:**

Fig10 marks four regions in the mass and mixing parameter space, which can explain the magnitude and energy dependence of the survival probability $P_{ee}$ shown in Table 1. They correspond to the so called Large Mixing Angle (LMA), Small Mixing Angle (SMA), Low Mass (LOW) and Vacuum Oscillation (VAC) solutions. For the LMA and SMA solutions ($\Delta m^2 \sim 10^{-5}$ eV$^2$) the low energy Ga experiment (E << 1 MeV) falls above the MSW triangle in $\Delta m^2/E$, while the SK and SNO experiments (E >>1 MeV) fall inside it. Therefore the solar matter effect can explain the observed decrease of the survival probability with increasing energy. For the LOW solution the low energy Ga experiment is pushed up to the region indicated by the dashed line, where it gets an additional contribution to the $P_{ee}$ from the earth's regeneration effect. Finally the VAC solution explains the energy dependence of the survival probability via the energy dependence of the oscillation phase in eq. (29). Fig 11 shows the predicted survival probabilities for the four solutions as functions of the neutrino energy [20]. The LMA and LOW solutions predict mild and monotonic energy dependence, while the SMA and VAC solutions predict very strong and nonmonotonic energy dependence.

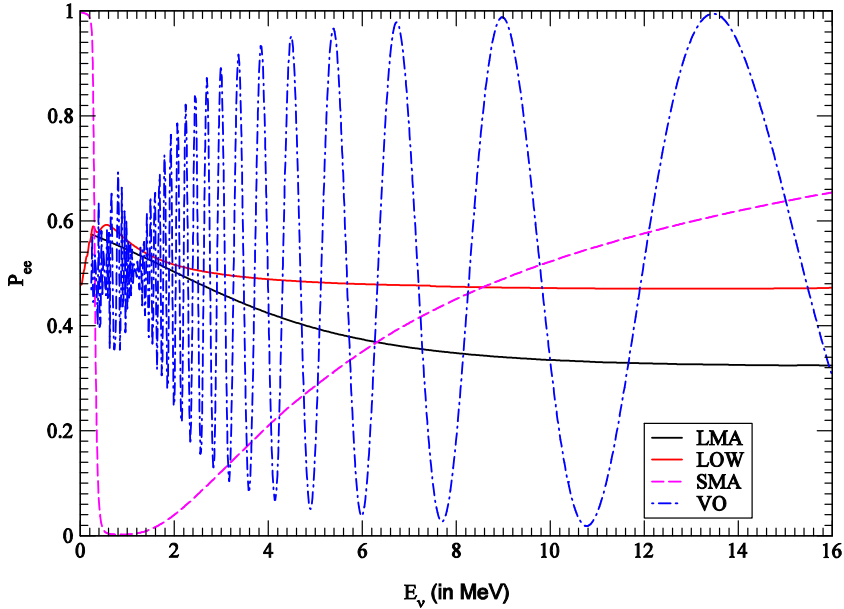

Fig.11.The predicted $\nu_e$ survival probabilities for the SMA, LMA, LOW and VAC solutions.

The survival rates in Table 1 show a slight preference for a nonmonotonic energy dependence, since the intermediate energy Chlorine experiment shows a little lower survival rate than SK. Therefore the SMA and the VAC solutions were the early favourites. However, the situation changed with the measurement of the energy spectrum by SK[16]. It shows practically no energy dependence in the 5-15 MeV range in clear disagreement with the SMA and the VAC predictions of Fig. 11. This was supported by the charged current data from SNO [17]. So the SMA and VAC solutions were ruled out in favour of the LMA and LOW. We also see from Fig. 11 that the LOW solution can not account for the entire drop of the survival probability with energy – from 0.55 to 0.35. But one could blame the low survival rate seen by the Cl, SK and SNO CC reactions partly on the large uncertainty in the Boron neutrino flux of the SSM (Fig. 8). This changed however with the publication of the neutral current data by SNO [17,18]. Being flavour independent, the NC reaction is unaffected by neutrino oscillation; and hence it can be used to measure the boron neutrino flux. The measured flux was in agreement with the SSM prediction and significantly more precise than the latter. Using this flux in a global fit to the solar neutrino data essentially ruled out LOW in favour of the LMA solution [21].

**Confirmation and Sharpening of the LMA Solution by KamLAND:**

Independent confirmation of the LMA solution came from the reactor antineutrino data of the long baseline KamLAND (KL) experiment [22]. It is a 1 kiloton liquid scintillator experiment detecting $\bar{\nu}_e$ from the Japanese nuclear reactors via the CC interaction $\bar{\nu}_e + p \to e^+ + n$, like the Cowan-Reines experiment [4], but with much greater efficiency and precision. It also measures the incident $\bar{\nu}_e$ energy via the visible scintillation energy produced by the positron and its annihilation with a target electron, i.e.

$$E_{vis} = E + m_e + m_p - m_n = E - 0.8 MeV. \tag{53}$$

The mean baseline distance of the detector from the reactors is $\langle l \rangle \sim 180$ km, which means it is sensitive to the $\Delta m^2 \geq 10^{-5}$ eV$^2$ region as mentioned earlier. Thus the experiment was designed to probe the LMA region. The first KL result from the 162 ty (ton-year) data showed a survival rate $P_{ee} \approx 0.6$. This was in perfect agreement with the LMA prediction , while effectively ruling out the alternative solutions. Moreover the observed spectral distortion of the KL data, taken together with the global solar neutrino data, narrowed down the $\Delta m^2$ to two sub-regions called low-LMA and high-LMA, corresponding to the 1$^{st}$ and 2$^{nd}$ oscillation nodes [22,23]. This was followed by the 766 ty KL data, which pinned down the mass parameter finally to the low-LMA band [24]. Fig 12 shows the energy spectrum of this KL data along with the no oscillation and the best oscillation solution fits [25]. The two fits come closest at $E_{vis} \approx 5$ MeV (E ≈ 6MeV), corresponding to the first oscillation node, $\langle l \rangle = \lambda = 180$ km. Substituting these values in eq. (30) gives $\Delta m^2 = 8 \times 10^{-5}$ eV$^2$.

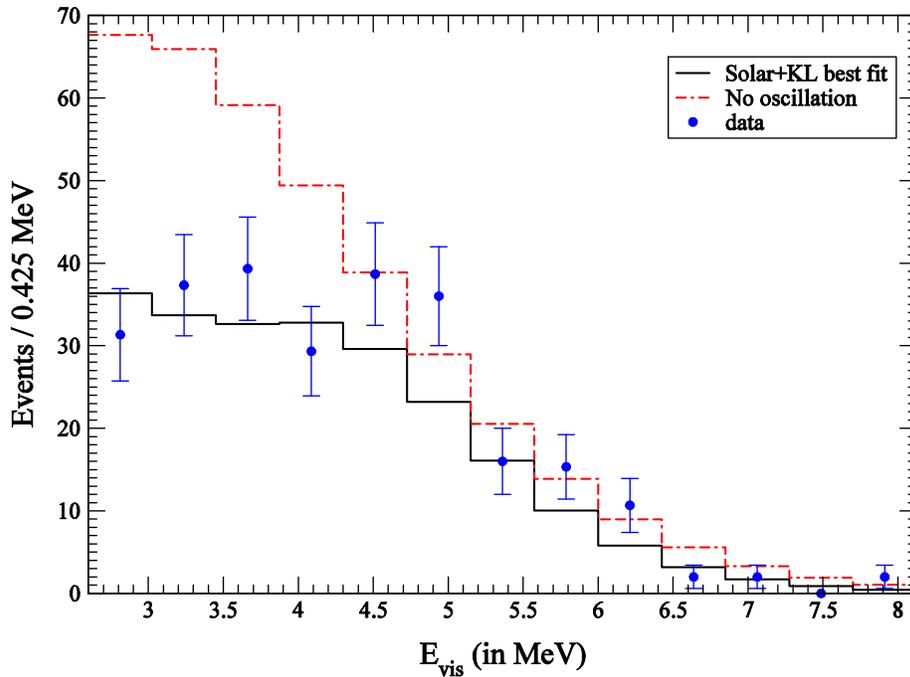

Fig.12. The 766 ty KL spectrum compared with the no-oscillation and the best oscillation fits [25].

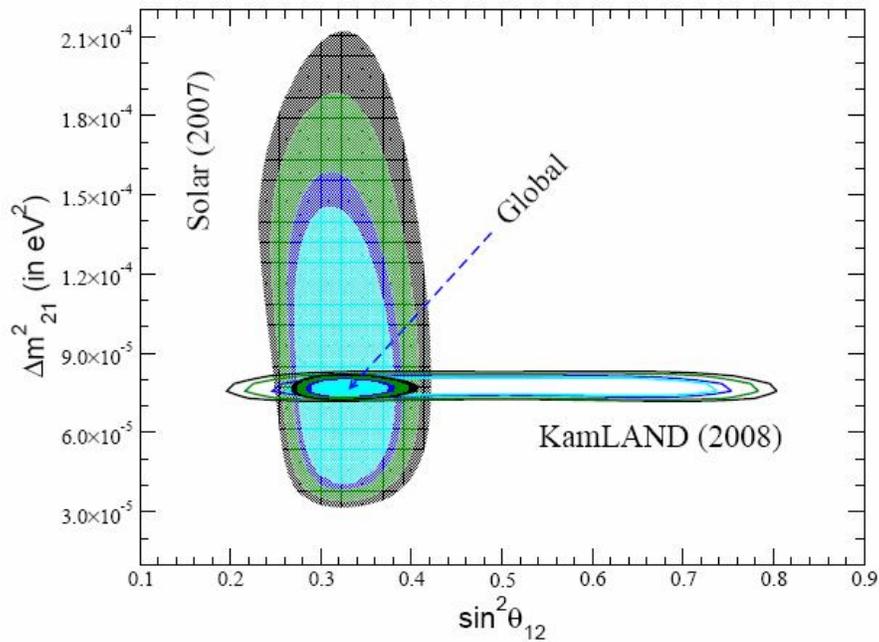

Fig.13. The 90, 95, 99, 99.73% Confidence Level fits to solar neutrino mass and mixing parameters from a global fit to solar neutrino and 2.8 kty KL data. The results of separate fits to these two data sets are also shown for comparison [27].

Very recently we have got the 2.8 kty KL data [26]. Fig. 13 shows the best fit values of the solar neutrino mass and mixing parameters, resulting from a global fit to solar neutrino and this high statistics KL data [27]. The results of separate fits to the solar and KL data are also shown for comparison. One sees that the KL and the solar neutrino data are primarily responsible for determining the mass and the mixing angle respectively. Thus combining them gives a precise estimate of both these oscillation parameters. The best fit values of these parameters are

$$\Delta m_{sol}^2 = \Delta m_{21}^2 = 7.7 \times 10^{-5} eV^2, \sin^2 \theta_{sol} = \sin^2 \theta_{12} = 0.33. \tag{54}$$

## 7. What Lies Ahead:

Evidently there has been a great advance in neutrino physics over the last decade. What used to be the mysteries of the missing solar and atmospheric neutrinos a decade back are now means of precise measurement of their mass and mixing parameters. It is therefore appropriate to take stock of the current status and future prospects of the neutrino mass and mixing measurements. Fig. 14 illustrates what we know now about the masses and mixings of the three light neutrinos from atmospheric and solar neutrino oscillation data along with those from accelerator and reactor neutrino experiments [28].

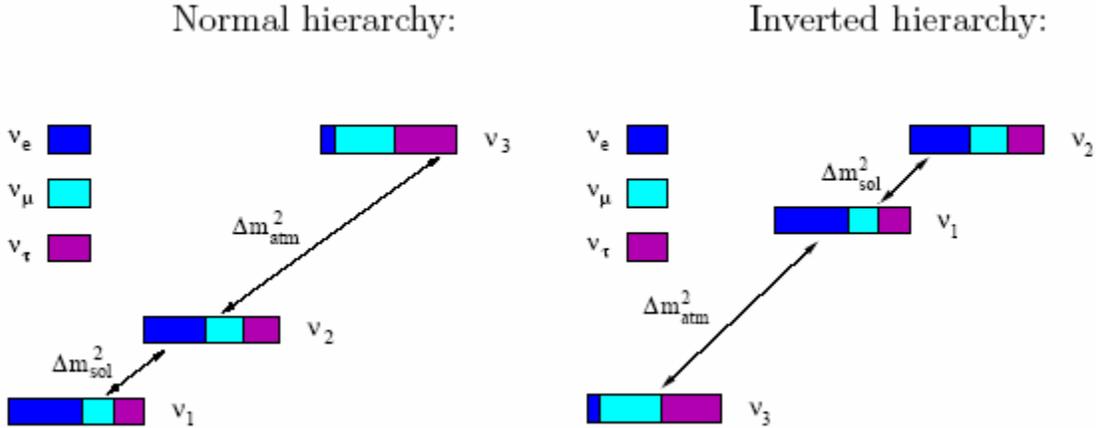

Fig.14. A schematic diagram of neutrino masses and mixings [28].

There are two mass-square gaps ($\Delta m^2$) between the three neutrino masses. Thanks to the solar matter effect, we know both the magnitude and sign of the smaller gap $\Delta m^2_{sol}$ (eq. 54), while we know only the magnitude of the larger gap $\Delta m^2_{atm}$ (eq. 37). We also know two of the three mixing angles, which are both large. These are the $\sin^2\theta_{sol}$ ($\approx 1/3$) of eq. (54), representing the $\nu_e$ component of $\nu_2$; and and the $\sin^2\theta_{atm}$ ($\approx 1/2$) of eq. (37), representing the $\nu_\tau$ component of $\nu_3$. However, we do not know the size of the third mixing angle, representing the $\nu_e$ component of $\nu_3$, but only its upper limit [29]

$$\sin^2 \theta_3 = \sin^2 \theta_{13} \leq 0.07 (\theta_{13} \leq 1/4). \tag{55}$$

The sign of the $\Delta m^2_{atm}$ leads to a two-fold ambiguity called normal and inverted hierarchy, as illustrated in the figure. Moreover, we do not know the absolute mass scale of the neutrinos – the lowest mass could be larger than the gaps, implying relatively large and degenerate neutrino masses. Finally, we do not know for sure the Majorana nature of these neutrino masses, since neutrino oscillation involves no L violation. Let me conclude by briefly discussing the prospects of determining each of these unknown entities in the foreseeable future (i.e. the next decade or two).

i) The third mixing angle $\theta_{13}$ represents the $\nu_e$ component involved in the atmospheric neutrino oscillation scale of eq. (36) ; i.e. $l \sim 10^3$km for E ~ GeV (accelerator) neutrinos or $l \sim$ km for E ~ MeV (reactor) neutrinos. The upper limit on $\sin^2\theta_{13}$ comes from the $\nu_e$ disappearance experiment at the CHOOZ nuclear reactor in France [29]. An improved version of this experiment called Double CHOOZ, using two detectors to cut down the systematic error, is expected to improve the detection limit of $\sin^2\theta_{13}$ by an order of magnitude. A similar experiment is also planned at the Daya Bay reactor in China. Moreover improved versions of the long baseline accelerator neutrino experiments of Japan [11] and USA [12] are planned with more intense $\nu_\mu$ beams and bigger detectors, called J2K and NOVA respectively. They can also improve the detection limit of $\sin^2\theta_{13}$ by an order of magnitude via $\nu_e$ appearance experiment.

ii) In principle the atmospheric neutrino experiment can determine the sign of $\Delta m^2_{atm}$ via the earth matter effect on the propagation of $\nu_e$, as mentioned in the context of solar neutrino oscillation. However, this is suppressed by the small $\nu_e$ component in atmospheric neutrino oscillation scale, i.e. $\sin^2\theta_{13}$. If this angle is found to be close to the upper limit of eq. (55), then one may be able to determine the sign of $\Delta m^2_{atm}$ from the atmospheric neutrino experiment at SK and the proposed India based neutrino observatory (INO). The INO detector is planned to use 50 kt of magnetized iron calorimeter, and has the advantage of distinguishing between $\nu_\mu$ and $\bar{\nu}_\mu$ interactions via the measurement of $\mu^\pm$ charge.

iii) The most robust limit on the absolute scale of the neutrino mass comes from a study of the end point spectrum of the tritium beta decay, which puts an upper limit on $\bar{\nu}_e$ mass [30], or rather the mass eigenstate dominated by $\bar{\nu}_e$, i.e.

$$m_{\nu_1} \leq 2eV. \qquad (56)$$

A much larger version of the tritium beta decay experiment called KATRIN, which is now under construction at Karlsruhe, Germany, will probe this mass down to 0.3 eV. Note however that this is still an order of magnitude higher than the mass gaps, i.e.

$$\sqrt{\Delta m^2_{atm}} \simeq 0.05 eV, \sqrt{\Delta m^2_{sol}} \simeq 0.01 eV.$$

iv) Finally, the Majorana nature of neutrino mass can be probed via second order weak nuclear beta decay processes called neutrino-less double beta decay (NDBD), i.e.

$$N(A,Z) \to N'(A, Z+2) + 2e^-, \tag{57}$$

which violates lepton conservation (ΔL=2). This will show up as a line signal in the di-electron energy spectrum against the continuum background from the lepton conserving process

$$N(A,Z) \to N'(A, Z+2) + 2e^- + 2\bar{\nu}_e. \tag{58}$$

Assuming Majorana nature of neutrino mass one has already got a strong limit of

$$m_{\nu_1} \leq 0.4 - 0.5 eV \tag{59}$$

from NDBD experiments with Germanium [31] and Tellurium [32]. Improved versions of these experiments are being planned, which will have an order of magnitude better detection limit than this. In that case one can hope to see a NDBD signal of Majorana mass for the inverted hierarchy scenario of fig. 14. For the normal hierarchy scenario however one would still be away from the goal.

In a more distant future one hopes also to probe CP violation in the lepton sector via neutrino oscillation. This is analogous to the observed CP violation effect in the quark sector due to their Yukawa interaction, for which Kobayashi and Maskawa got the 2008 Nobel prize in physics.

**Acknowledgement :** This work was supported in part by the National Initiative in Undergraduate Science (NIUS) programme of HBCSE and in part by the DAE (BRNS) through the Raja Ramanna Fellowship. I thank my NIUS students and in particular Ipsita Satpathy for their feedback. I am also thankful to Profs Subhendra Mohanty and Utpal Sarkar for discussions.